# Sustainable and Efficient Renewable-Driven Energy Trading via Neural-Enhanced Time-Adaptive Robust Nash Bargaining between Hydrogen-Enriched Gas and Active Distribution Networks

Wenwen Zhang

***Abstract*—Integrated hydrogen-enriched compressed natural gas (HCNG) and active distribution network (ADN) is providing efficient and sustainable flexibility for consuming renewable energies. Yet, cross-sector privacy and uncertain high-renewable scenarios block stable coordination. They also worsen decision performance and convergence. To conquer the barrier, a neural-enhanced time-adaptive robust Nash bargaining strategy is proposed. In the first stage, to clear energy trading between ADN and gas distribution network (GDN) and promote its sustainability, a privacy-preserved Nash Bargaining based on the alternating direction method of multipliers (ADMM) is applied. The next robust dispatch stage explores the worst renewable scenarios and de-risks ADN's profit collapse from uncertainties. The convergence of the entire energy trading scheme is theoretically proved. As such, sustainable returns from the participation of solid oxide fuel cell (SOFC) and HCNG are facilitated. Finally, a time complexity and social welfare co-driven neural-enhanced time-adaptive strategy is proposed. The strategy assesses the influence of time resolution on social benefits and solving time in multi-energy trading. Based on the assessment, a neural network surrogate model is trained to accelerate the trading process in a close-looped manner. Numerical assessment reveals that, the proposed strategy reaps a stable social welfare of nearly 1.6% to total cost, and benefit-steady situations for both ADN and GDN, even in the worst renewable scenarios. Moreover, it reduces runtime to 102.47s, improving computational efficiency by over 69.86% versus the fixed time-scale baseline, almost without sacrifice in economy.**

***Index Terms*—Hydrogen enriched compressed natural gas, two-stage robust Nash bargaining, solid oxide fuel cells, hybrid energy storage system, time-adaptive, neural network, sustainable energy scenarios**

## NOMENCLATURE

*A. Abbreviation*

| | |
|---|---|
| HCNG | Hydrogen-enriched compressed natural gas |
| IENGS | Integrated electricity and natural gas systems |
| ADN | Active distribution network |
| GDN | Gas distribution network |
| HGN | Higher-level gas network |
| P2G | Power to gas |
| G2P | Gas to power |
| DER | Distributed energy resource |
| ET | Electrolytic tank |
| HT | Hydrogen tank |
| SOFC | Solid oxide fuel cell |
| TG | Transmission grid |
| HESS | Hybrid energy storage system |
| C&CG | Column-and-constraint generation |
| ADMM | Alternating direction method of multipliers |
| BCD | Block coordinate descending |

*B. Sets/ Indices*

| | |
|---|---|
| $\mathbb{J}_1/\mathbb{J}_2/\mathbb{J}_3/\mathbb{J}_4$ | Set of P2Gs/ G2Ps/li-ion batteries/DERs |
| $\mathbb{A}/\mathbb{B}$ | Set of nodes in GDN/ ADN |
| $\mathbb{T}$ | Set of operation periods |
| $m/n/g$ | Index of natural gas nodes |
| $i/j/h$ | Index of power nodes |
| $j_1/j_2/j_3/j_4$ | Index of P2G/ G2P/ li-ion battery/ DER |
| $t$ | Index of operation periods |

*C. Parameters*

| | |
|---|---|
| $\Delta t$ | Time interval |
| $\varepsilon_t/\mu_t$ | Prices of natural gas/ electricity |
| $\beta^{ET}/\beta^{SOFC}$ | Power cost of ET/ SOFC |
| $\beta^{Li}/\alpha^{Li}$ | Power/ capacity cost of li-ion battery |
| $\alpha^{HT}/V^{HT}$ | Capacity cost/volume of HT |
| $\xi_{j_3}^{Li.min}/\xi_{j_3}^{Li.max}$ | Minimum/maximum charge status |
| $w_m^{min}/w_m^{max}$ | Minimum/ maximum node pressure |
| $\omega$ | Max volume fraction of hydrogen |
| $\phi^{LD}/\phi^{DER}$ | Volatility range of uncertainties |
| $\eta^{P2G}/\eta^{P2G}$ | Conversion efficiency of P2G/G2P |
| $c_{mn}$ | Weymouth constant of the pipeline |
| $C^{G}/C^{E}$ | Daily operating costs of GDN/ADN |
| $C^{HGN}$ | Purchase costs of Natural gas from HGN |
| $C^{P2G}$ | Purchase costs of gas from P2G |
| $C^{G2P}/C^{TG}$ | Purchase costs of electricity from G2P/ upstream transmission grid |
| $C^{Li}/C^{SOFC}/C^{ET}/C^{HT}$ | Usage costs of battery/ SOFC/ ET/ HT |
| $HHV^{CH_4}/HHV^{H_2}$ | High heat value of CH4/ H2 |
| $P_{j_1}^{ET}/P_{j_2}^{SOFC}/P_{j_3}^{Li}$ | Rated power of ET/ SOFC/ li-ion battery |
| $E_{j_3}^{Li}$ | Rated capacity of li-ion battery |
| $T^{ET}/T^{SOFC}/N_{j_3}^{Li}$ | Lifetime of ET/SOFC/ li-ion battery |
| $U_i^{min}/U_i^{max}$ | Min/max node voltage |
| $r_{ij}/x_{ij}$ | Branch resistance/ reactance |

*D. Variables*

| | |
|---|---|
| $\vartheta_t^{P2G}/\vartheta_t^{G2P}$ | Price of power for P2G/ gas for G2P |
| $P_{j_1,t}^{P2G}/P_{j_2,t}^{G2P}$ | Power from P2G/G2P |
| $P_{j_3,t}^{Li.disc}/P_{j_3,t}^{Li.c}$ | Li-ion battery discharge /charge power |
| $E_{j_3,t}^{Li}$ | Li-ion battery store energy |
| $D_{j_3}/\xi_{j_3,t}^{Li}$ | Li-ion battery charge depth/ charge status |
| $P_{j_4,t}^{DER}/Q_{j_4,t}^{DER}$ | Active/ reactive power injected by DER |
| $P_t^{TG}$ | Electricity purchased from TG |
| $P_{j,t}^{LD}/Q_{j,t}^{LD}$ | Active/ reactive power load |
| $I_{ij,t}/U_{i,t}$ | Branch current/ Node voltage |
| $P_{ij,t}/P_{jh,t}/$ | Active power flow |
| $Q_{ij,t}/Q_{jh,t}$ | Reactive power flow |
| $G_{j_2,t}^{G2P}$ | Volume of gas from G2P |
| $G_{n,t}$ | Equivalent gas load volume |
| $G_{mn,t}/G_{ng,t}$ | Branch gas flow |
| $G_t^{HGN}$ | Natural gas purchased from HGN |
| $G_{n,t}^{H_2}/G_{n,t}^{LD}$ | Volume of hydrogen injection/ gas load |
| $G_{j_1,t}^{HT}$ | Hydrogen volume stored in HT |
| $w_{m,t}/w_{n,t}$ | Node pressure |
| $G_t^{CH_4}/G_t^{H_2}$ | CH4/ H2 injected into the pipeline |
| $P_{j,t}^{LD.pre}/P_{j_4,t}^{DER.pre}$ | Power load/ DER output forecast value |
| $P_{j,t}^{LD.real}/P_{j_4,t}^{DER.real}$ | Power load/ DER output actual value |
| $P_{j_3,t}^{Li.pre}$ | Power of li-ion battery in the first stage |
| $\Delta P_{j_3,t}^{Li}$ | Power adjustment value of li-ion battery |

## I. INTRODUCTION

While the penetration of distributed energy resources (DERs) into ADNs is crucial for decarbonization [1,2], their inherent stochasticity causes generation-load mismatches, undermining DER utilization and network stability [3]. Regarding this issue, recent research proposes the formation of an integrated electricity and natural gas systems (IENGS) by combining the ADN and GDN in sustainable energy scenarios[4], which have emerged as the flexible way to manage uncertain DERs and promote energy efficacy [5], drawing growing studies worldwide.

To further unleash the flexibility and efficiency of IENGS in [4,5], [6] proposes injecting hydrogen into the GDN to form an HCNG-based IENGS, and verifies the advantages of the proposed IENGS in local balancing of DERs and minimizing operational costs. Furthermore, the practical viability of HCNG is further corroborated by the French "GRYHD" project [7], which confirms its benefits of lower emissions and higher generation efficiency. Additionally, compared to the gas turbines in the IENGS mentioned in [6], SOFCs have been demonstrated the better electrical efficiency and the wider fuel adaptability [8,9]. Existing literature also revealed that, oriented to power grid steady operation applications, SOFC can be generically modelled following the gas turbine's steady states [10]. Moreover, [11] validates the system stability of introducing SOFCs into ADN. Obviously, the above research elucidates the feasibility and enormous development prospects of the HCNG-enabled IENGS bridged with SOFCs, however, studies on energy dispatch for such systems are still scarce.

Upon the highly efficient infrastructure of HCNG-enabled GDN and SOFCs, profit pattern that advances development of the IENGS becomes another imminent issue. Prior research confirms that centrally controlled IENGS enhance stochasticity mitigation, renewable consumption, and social welfare [12]. More cognitions have manifested that centralized IENGSs are away from engineering, where GDN and ADN are usually operated by different independent entities. Even though these entities partially privatize their information, [13] proved that IENGS still benefits rich social welfare via the ADMM. A deeper issue in such cooperative markets is that policy biases (e.g., penalties for renewable curtailment) can weaken one stakeholder, hindering optimal social welfare [14]. Proper benefit sharing strategies are the key to solve the issue. Nash bargaining theory, proven in other energy contexts, offers a promising solution for its dual benefits of social welfare maximization and fair distribution [15,16]. Nevertheless, its application to ADN-GDN cooperation is hampered by a high-dimensional variable space that im-pedes ADMM convergence. Current solutions include algorithmic enhancements to ADMM [17,18] and model simplification via adaptive time-windows [19,20] extends [19] by developing a closed-loop time adaptive scheduling model with neural network-based performance feedback. However, its availability in the IENGS has not been investigated, let alone the concerned IENGS, where GDN dominates the cooperation due to efficient HCNG-fueled SOFC and HCNG network.

Despite benefit sharing can facilitate stable operation of IENGS, profit collapse may still exist in the ADN side. Under the incentives of high-profit latency, SOFCs and HCNG network may participate more in ADN. Their slow responses force the ADN to buy costly electricity from upstream transmission grid to balance uncertainties of DERs [21,22], triggering noticeable profit loss especially when there exists transactions bias due to over-prediction of DERs. [23] suggested configuring hybrid energy storage system (HESS) to enhance rapid power balancing capability, and making look-ahead dispatch based on robust optimization to withstand slow-response hydrogen-to-power sources. Similar idea is also verified in [24,25], where operational risks from uncertainties of renewable energy in traditional IENGS are mitigated by two-stage robust model. But the foregoing studies were not embarked on more complicated benefit sharing for standalone GDN and ADN. Furthermore, regarding the solution of the two-stage robust optimization issue, [26] inspired to use block coordinate descending (BCD) algorithm to ease the intact two-stage model into tractable tri-level equivalence. However, the two-stage robust mechanism based on BCD has yet to be validated for multi-entity IENGS coordinated via Nash bargaining, especially when neural network feedback is applied to improve ADMM convergence through adaptive time-window optimization.

This paper brought forward configuration of multi-entity IENGS, benefit sharing mechanism to promote win-win colabration, and tailored algorithm to dispatch the whole system. The main contributions are three-fold:

(1) In high-penetration sustainable energy scenarios, a novel IENGS model bridges the HCNG and ADN via SOFC. This setup establishes an electricity- HCNG -electricity cycle, unlocking cooperative potential. Specifically, it allows HCNG production to bypass the methanation process, reducing energy dissipation by 17%. Further, SOFCs achieve an energy conversion cost saving of $0.04/kWh compared to gas turbines.

(2) To counter the risk of abnormally negative electrolytic hydrogen prices, driven by renewable curtailment penalties and the high efficiency of HCNG collaboration, a Nash bargaining-based benefit-sharing mechanism is proposed. Built upon ADMM algorithm, this mechanism safeguards participant privacy and fosters a stable, incentive-compatible partnership.

(3) For the first time, a time-adaptive two-stage robust Nash bargaining strategy with binding batteries is proposed towards more economical and stable IENGS. The first stage places the proposed Nash bargaining between ADN and GDN, and the second stage leverages robust dispatch of li-ion batteries to promote economic benefit of ADN. Besides, a novel algorithm combining ADMM and BCD, bridged by column-and-constraint generation (C&CG), is developed to solve this complex model efficiently, with its convergence proven in the appendix.

## II. Framework of The Proposed HCGN-enabled IENGS

Fig. 1 illustrates the neural network-enhanced time-adaptive two-stage robust cooperative dispatch strategy for the HCNG enabled GDN and ADN. The process begins with k-means clustering of historical data to define adaptive time windows, re-placing fixed one-hour intervals.

The first stage formulates the cooperation as a Nash bargaining game, determining optimal energy trading volumes and prices between the entities via power-to-hydrogen and SOFC infrastructures to maximize and fairly share social welfare. The second stage uses these trading baselines to conduct robust dispatch of li-ion batteries, helping the ADN manage DER uncertainty. This involves identifying worst-case DER scenarios and then optimizing battery schedules to minimize grid purchases and correct transaction deviations. The details are provided in Section III and Section IV. Then, the above two-stage optimization is solved iteratively under various window settings. The performance data forms a training set for a neural network. The trained encoder then provides the prerequisite optimal temporal resolution by balancing both social benefit deviation and computational time, as detailed in Section V.

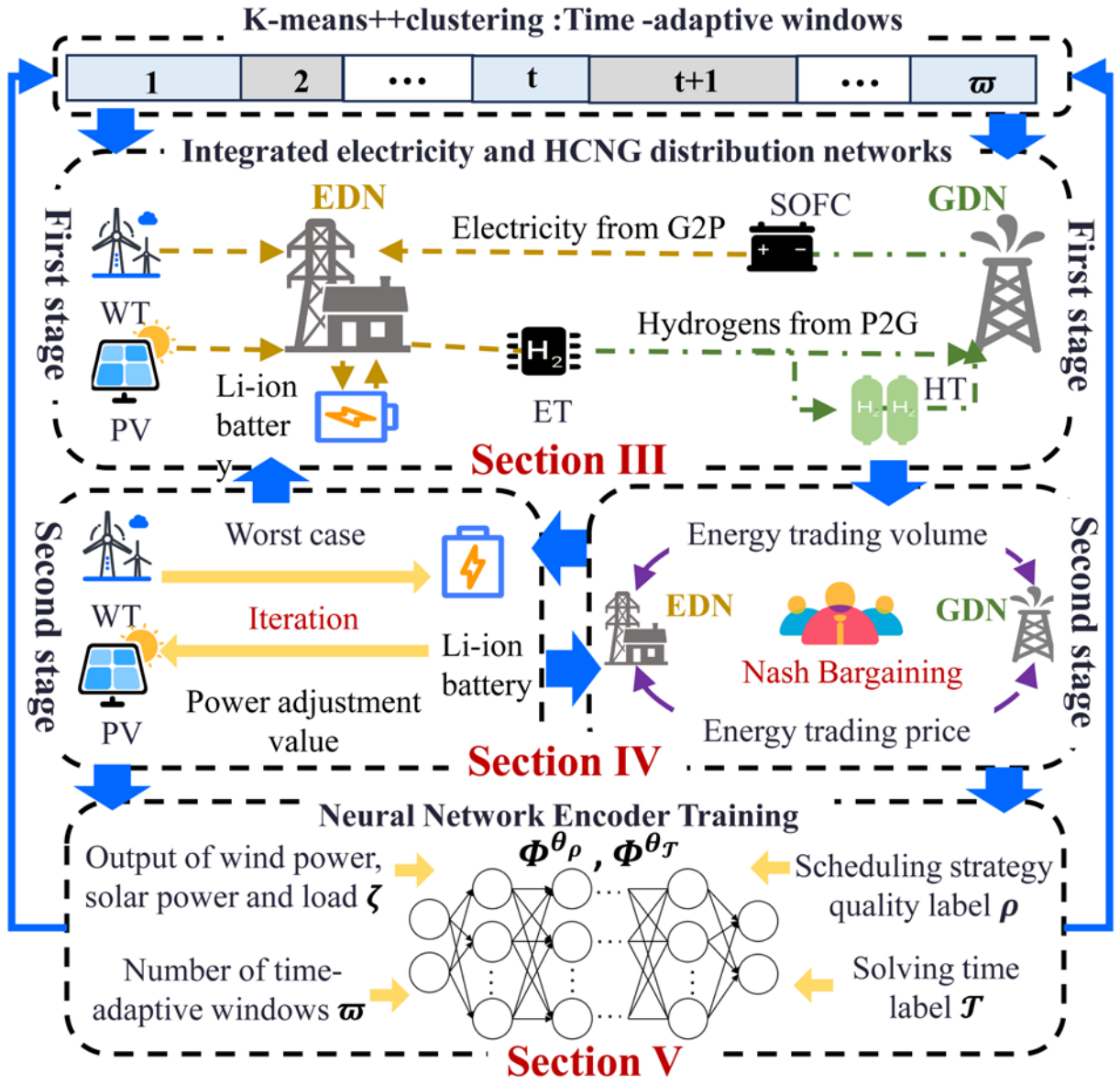

**Fig. 1.** The proposed neural network-enhanced time-adaptive two-stage robust benefit sharing framework.

## III. Time-Adaptive Scheduling Model of Integrated Electricity and HCNG Distribution Networks

Following the designated NCGN-enabled IENGS, we settle the cooperative scheduling model for both utilities of HCNG-enabled GDN (i.e., the HCNG network) and ADN. Each entity works to increase its own profit, and is allowed to interact with the other entity via P2G or G2P equipment for more benefits. To improve energy conversion efficiency and promote latent social welfare, we bring in highly efficient P2G facility of HCNG-enabled SOFC, and a compact HCNG network that can bypass additional conversion from hydrogen to methane. Furthermore, the cooperative scheduling model integrates adaptive time windows, effectively balancing scheduling accuracy and computational complexity for IENGS.

### *A. Time Aggregation Based on Robust K-means++*

To establish the adaptive time resolution set for energy exchange scheduling between GDN and ADN, the hourly time series $\boldsymbol{\zeta}$ of wind power, solar power and load are clustered via the robust *k-means*++. Details are provided in (1)-(2):

$$\boldsymbol{\zeta} \xrightarrow{\text{Robust } k\text{-means++}} \tilde{\boldsymbol{\zeta}} \triangleq [\boldsymbol{\kappa}_1, \dots, \boldsymbol{\kappa}_t, \dots, \boldsymbol{\kappa}_\varpi] \in \mathbb{R}^{3\times\varpi} \tag{1}$$

$$\Delta t_t \triangleq N_t \Delta, \Delta\mathbf{T} = [\Delta t_1, \dots, \Delta t_t, \dots, \Delta t_\varpi]^{\mathrm{T}} \in \mathbb{R}^{\varpi\times 1} \tag{2}$$

where $\tilde{\boldsymbol{\zeta}}$ is the set of clustering centers derived from sequence $\tilde{\boldsymbol{\zeta}}$, $\boldsymbol{\kappa}_t$ represents the aggregated renewable generation (wind/solar) and load demand in the *t*-th adaptive time window. $\varpi$ is the number of time-adaptive windows. $\Delta t_t$denote the duration of the *t*-th adaptive time window, which depends on the number ($N_t$) of time durations ($\Delta$) included in the *t*-th cluster, $\Delta$ represents 1 hour.

### *B. Scheduling Model of HCNG-enabled Network*

The entity of the HCNG-enabled GDN intends to optimize its own yields, such that its own scheduling model will be set to maximize the income of selling gas to ADN, to minimize purchase cost of natural gas from higher-level gas network (HGN), purchase costs of hydrogen from P2G, and use costs of electrolytic tank (ET)/ hydrogen tank (HT), and to adhere to operational constraints of HCNG network, as given below:

$$\min_{P_{j_1,t}^{P2G}, G_{j_2,t}^{G2P}, G_{j_1,t}^{HT}, \vartheta_t^{P2G}, \vartheta_t^{G2P}} C^G \tag{3}$$

$$\begin{cases} C^G = C^{HGN} + C^{P2G} + C^{ET} + C^{HT} - C^{G2P} \\ C^{HGN} = \sum_{t=1}^{\varpi} \varepsilon_t G_t^{HGN}, C^{ET} = \sum_{j_1} \sum_{t=1}^{\varpi} \beta^{ET} P_{j_1,t}^{P2G} \Delta t_t / P_{j_1}^{ET} T^{ET} \\ C^{P2G} = \sum_j \sum_{t=1}^{\varpi} \vartheta_t^{P2G} P_{j_1,t}^{P2G} \Delta t_t \\ C^{G2P} = \sum_{j_2} \sum_{t=1}^{\varpi} \vartheta_t^{G2P} G_{j_2,t}^{G2P}, C^{HT} = \sum_{j_1} \alpha^{HT} V_{j_1}^{HT} / T^{HT} \end{cases} \tag{4}$$

$$G_{n,t} = \sum_m G_{mn,t} + G_{n,t}^{H_2} + G_{n,t}^{HGN} - G_{n,t}^{G2P} - \sum_g G_{ng,t} \tag{5}$$

$$G_{j_1,t}^{H_2} = G_{j_1,t}^{P2G} - G_{j_1,t}^{HT}, G_{j_1,t}^{P2G} = \eta^{P2G} \cdot P_{j_1,t}^{P2G} . \Delta t_t / HHV^{H_2} \tag{6}$$

$$G_{n,t} = (G_{n,t}^{LD} HHV^{CH_4}) / HHV^{mix}, G_t^{H_2} / (G_t^{H_2} + G_t^{CH_4}) \le \omega \tag{7}$$

$$HHV^{mix} = \omega HHV^{H_2} + (1-\omega) HHV^{CH_4} \tag{8}$$

$$V_{j_1,t}^{HT} = V_{j_1,t-1}^{HT} + G_{j_1,t}^{HT}, 0 \le V_{j_1,t}^{HT} \le V_{j_1}^{HT} \tag{9}$$

where the balance constraints of the gas network nodes are presented as (5)-(6). Changes in gas composition can affect the consumed volume [27], equivalent replacement of gas loads as shown in (7)-(8). (9) is the operational constraint of HT.

$$G_{mn,t} = \tau_{mn,t} \cdot c_{mn} \sqrt{\tau_{mn,t}\left({w_{m,t}}^2 - {w_{n,t}}^2\right)} \tag{10}$$

$$\tau_{mn,t} = \begin{cases} 1 & w_{m,t} \ge w_{n,t} \\ -1 & w_{m,t} \le w_{n,t} \end{cases} \tag{11}$$

where the steady-state natural gas power flow constraints are formulated as (10)-(11) upon Weymouth equation. Considering that the GDN is radiant, the second-order cone relaxation can be used to convexify the constraints (10)-(11) into (12) [28].

$$\left\| \begin{matrix} G_{mn,t} \\ c_{mn} w_{n,t} \end{matrix} \right\|_2 \le c_{mn} w_{m,t}, w_m^{min} \le w_{m,t} \le w_m^{max} \tag{12}$$

To ensure that the second-order cone relaxation of natural gas power flow is exact, a penalty term on the nodal pressure difference is added to the objective function (13) [29].

$$\min_{P_{j_1,t}^{P2G}, G_{j_2,t}^{G2P}, G_{j_1,t}^{HT}, \vartheta_t^{P2G}, \vartheta_t^{G2P}} \left( C^G + \sum_t^{\varpi} \sum_{m,n} \tau\left(w_{m,t} - w_{n,t}\right)\right) \tag{13}$$

$$\forall t \in \mathbb{T} = \{1,2, \dots \varpi\}, j_1 \in \mathbb{J}_1, j_2 \in \mathbb{J}_2, m/n/g \in \mathbb{A}$$

*C. Scheduling Model of Electric Distribution Network*

The optimization goal of the ADN is to minimize daily operating costs in the worst scenarios, including purchase costs of electricity from upstream transmission grid, purchase costs of electricity from G2P units, and HESS usage costs, and to maximize incomes from the sale of P2G gas, as shown in (14)-(15).

$$\min_{P_{j_3,t}^{Li}, P_{j_1,t}^{P2G}, P_{j_2,t}^{G2P}, \vartheta_t^{P2G}, \vartheta_t^{G2P}} C^E \tag{14}$$

$$\begin{cases} C^E = C^{G2P} + C^{TG} + C^{HESS} - C^{P2G} \\ C^{TG} = \sum_{t=1}^{\varpi} \mu_t P_t^{TG} \Delta t_t, C^{HESS} = C^{Li} + C^{SOFC} \\ C^{Li} = \sum_{j_3} \sum_{t=1}^{\varpi} \left(\alpha^{Li} E_{j_3}^{Li} + \beta^{Li} P_{j_3}^{Li}\right) \cdot \left(P_{j_3,t}^{Li.disc} - P_{j_3,t}^{Li.c}\right) \cdot \Delta t_t / \left(N_{j_3}^{Li} \cdot E_{j_3}^{Li} \cdot D_{j_3}\right) \\ N_{j_3}^{Li} = a_1 e^{(b_1 D_{j_3})} + a_2 e^{(b_2 D_{j_3})} \\ C^{SOFC} = \sum_{j_2} \sum_{t=1}^{\varpi} \beta^{SOFC} P_{j_2,t}^{G2P} \Delta t_t / \left(P_{j_2}^{SOFC} T^{SOFC}\right) \end{cases} \tag{15}$$

where HESS usage costs consist of li-ion battery usage costs $C^{Li}$ and SOFC usage costs $C^{SOFC}$. $C^{Li}$ considers that depth of discharge (DOD) can affect the cycle life of li-ion battery [30], $N_{j_3}^{Li}$ is the number of cycles when the DOD is $D_{j_3}$, and $a_1$, $a_2$, $b_1$, $b_2$ are correlation coefficients

Distribution network power flow is formed as a standard second-order cone as in (16)-(17):

$$\hat{I}_{ij,t} = I_{ij,t}^2, \hat{U}_{i,t} = U_{i,t}^2 \tag{16}$$

$$\left\| \begin{matrix} 2P_{ij,t} \\ 2Q_{ij,t} \\ \hat{I}_{ij,t} - \hat{U}_{i,t} \end{matrix} \right\|_2 \le \hat{I}_{ij,t} + \hat{U}_{i,t} \tag{17}$$

$$\left(U_i^{min}\right)^2 \le \hat{U}_{i,t} \le \left(U_i^{max}\right)^2 \tag{18}$$

$$P_{j,t}^{LD} = \sum_i P_{ij,t} - r_{ij} \hat{I}_{ij,t} - \sum_h P_{jh,t} + P_{j,t}^{DER} + P_{j,t}^{G2P} + P_{j,t}^{Li} - P_{j,t}^{P2G} \tag{19}$$

$$P_{j_2,t}^{G2P} = G_{j_2,t}^{G2P} \cdot HHV_t^{mix} . \eta^{G2P} \tag{20}$$

$$P_{j_3,t}^{Li} = P_{j_3,t}^{Li.pre} = P_{j_3,t}^{Li.disc} + P_{j_3,t}^{Li.c} \tag{21}$$

$$Q_{j,t}^{LD} = \sum_i Q_{ij,t} - x_{ij} \hat{I}_{ij,t} - \sum_h Q_{jh,t} + Q_{j,t}^{DER} \tag{22}$$

where (18)-(22) show the voltage security and power balance constraints for each bus in ADN.

$$E_{j_3,t}^{Li} = E_{j_3,t-1}^{Li} + \int_{t-1}^{t} P_{j_3,t}^{Li} \Delta t_t \tag{23}$$

$$\xi_{j_3,t}^{Li} = E_{j_3,t}^{Li} / E_{j_3}^{Li}, \xi_{j_3}^{Li.min} \le \xi_{j_3,t}^{Li} \le \xi_{j_3}^{Li.max} \tag{24}$$

$$\begin{cases} -P_{j_3}^{Li} \le P_{j_3,t}^{Li.c} \le 0, 0 \le P_{j_3,t}^{Li.disc} \le P_{j_3}^{Li} \\ 0 \le P_{j_2,t}^{G2P} \le P_{j_2}^{SOFC}, 0 \le P_{j_1,t}^{P2G} \le P_{j_1}^{ET} \end{cases} \tag{25}$$

$$\forall t \in \mathbb{T} = \{1,2, \dots, \varpi\}, j_1 \in \mathbb{J}_1, j_2 \in \mathbb{J}_2, j_3 \in \mathbb{J}_3, i,j,h \in \mathbb{B}$$

where (23)-(25) are the operational constraints of li-ion batteries, SOFCs, ETs.

## IV. Two-stage Robust Nash Bargaining-based Benefit Sharing Between HCNG Network and ADN

To actualize the social welfare of the IENGS set on the HCNG network and the ADN, an all-win cooperation is imperative. Otherwise, ADN will be placed in a weak position, since it is stressed to consume RES, but HCNG network features requisite energy storage property to this end. To guarantee the benefits for the participating utilities, this section firstly introduces a privacy-preserving benefit sharing mechanism based on Nash bargaining theory and ADMM. On the other hand, motivated by the salient energy conversion efficiency of SOFCs and HCNG network, their foreseeable aggressive participation can jeopardize the ADN's ability to follow the shift of DER. Whereon, to de-risk reciprocity dis-equilibration between the two entities, a robust scheduling aided by li-ion batteries will be drawn into the whole cooperative model. Packed the above built models together via C&CG, we finally settle a two-stage robust Nash bargaining model aimed at stable social welfare.

*A. Nash Bargaining between HCNG Network and ADN*

The interaction between the GDN and the ADN is modelled using cooperative game, and the Pareto optimum between the two entities can be achieved fairly by constructing the following Nash bargaining model:

$$\begin{cases} \max\limits_{P_{j_1,t}^{P2G},P_{j_2,t}^{G2P},\vartheta_t^{P2G},\vartheta_t^{G2P}} (C_0^E - C^E)(C_0^G - C^G) \\ C^E \le C_0^E,\, C^G \le C_0^G,\, s.t.\,(3)\text{-}(13),(16)\text{-}(25) \end{cases} \tag{26}$$
$$\forall t \in \mathbb{T} = \{1,2,\dots,\varpi\},\ j_1 \in \mathbb{J}_1,\ j_2 \in \mathbb{J}_2$$

where $C_0^E$, $C_0^G$ are the operating costs of ADN and GDN when they operate independently.

To solve the problem (26), we equivalently transform (26) into total benefit maximization problem (**Question 1**). The transaction price variables ($[\vartheta_t^{P2G},\vartheta_t^{G2P}]$) in **Question 1** are eliminated, thus the volume of energy transaction ($[P_{j_1,t}^{P2G},G_{j_2,t}^{G2P}]$) in **Question 1** can be directly solved to obtain the solution $[\check{P}_{j_1,t}^{P2G},\check{G}_{j_2,t}^{G2P}]$.

**Question 1:**

$$\begin{cases} \max\limits_{P_{j_1,t}^{P2G},P_{j_2,t}^{G2P}} (C_0^E - C^{TG} - C^{HESS}) + (C_0^G - C^{HGN} - C^{ET}) \\ s.t.\ (3)\text{-}(13),(15)\text{-}(25), \end{cases} \tag{27}$$
$$\forall t \in \mathbb{T} = \{1,2,\dots,\varpi\},\ j_1 \in \mathbb{J}_1,\ j_2 \in \mathbb{J}_2$$

By converting (26) into logarithmic form, we obtain another equivalent optimization problem (**Question 2**).

**Question 2:**

$$\begin{cases} \max\limits_{\vartheta_t^{P2G},\vartheta_t^{G2P}} \ln(C_0^E - C^E) + \ln(C_0^G - C^G) \\ C^E \le C_0^E,\, C^G \le C_0^G,\, s.t.\,(3)\text{-}(13),(15)\text{-}(25), \forall t \in \mathbb{T} \end{cases} \tag{28}$$

where decision variables $[P_{j_1,t}^{P2G},G_{j_2,t}^{G2P}]$ are replaced with fixed parameter $[\check{P}_{j_1,t}^{P2G},\check{G}_{j_2,t}^{G2P}]$. Whereupon optimal energy transaction prices $[\check{\vartheta}_t^{P2G},\check{\vartheta}_t^{G2P}]$ can be obtained by solving (28).

To ensure the privacy of the internal information of each entity, distributed optimization formulation (30)-(33) of (27)-(28) are built based on ADMM algorithm. Besides, to decouple $[P_{j_1,t}^{P2G},G_{j_2,t}^{G2P}]$ and $[\vartheta_t^{P2G},\vartheta_t^{G2P}]$, auxiliary variables are introduced in constraint (29) which needs to be enforced in the later.

$$\begin{cases} P_{j_1,t}^{P2G_G} = P_{j_1,t}^{P2G_P},\, G_{j_2,t}^{G2P_G} = G_{j_2,t}^{G2P_P} \\ \vartheta_t^{P2G_G} = \vartheta_t^{P2G_P},\, \vartheta_t^{G2P_G} = \vartheta_t^{G2P_P} \\ \forall t \in \mathbb{T} = \{1,2,\dots,\varpi\},\ j_1 \in \mathbb{J}_1,\, j_2 \in \mathbb{J}_2 \end{cases} \tag{29}$$

where $[P_{j_1,t}^{P2G_G},G_{j_2,t}^{G2P_G},\vartheta_t^{G2P_G},\vartheta_t^{P2G_G}]$ represents the GDN's willingness to trade, $[P_{j_1,t}^{P2G_P},G_{j_2,t}^{G2P_P},\vartheta_t^{G2P_P},\vartheta_t^{P2G_P}]$ represents the ADN's willingness to trade.

**Question 1:**

$$\begin{cases} \min L_1^E = C^{TG} + C^{HESS} - C_0^E + \\ \sum_{j_1}\sum_t^{\varpi} \chi_{j_1,t}^P (P_{j_1,t}^{P2G_G} - P_{j_1,t}^{P2G_P}) + \frac{\rho_1}{2}(P_{j_1,t}^{P2G_G} - P_{j_1,t}^{P2G_P})^2 + \\ \sum_{j_2}\sum_t^{\varpi} \chi_{j_2,t}^G (G_{j_2,t}^{G2P_G} - G_{j_2,t}^{G2P_P}) + \frac{\rho_2}{2}(G_{j_2,t}^{G2P_G} - G_{j_2,t}^{G2P_P})^2 \\ s.t.\,(16)\text{-}(25), \forall t \in \mathbb{T} = \{1,2,\dots,\varpi\},\ j_1 \in \mathbb{J}_1,\, j_2 \in \mathbb{J}_2 \end{cases} \tag{30}$$

$$\begin{cases} \min L_1^G = C^{HGN} + C^{ET} - C_0^G + \\ \sum_{j_1}\sum_t^{\varpi} \chi_{j_1,t}^P (P_{j_1,t}^{P2G_G} - P_{j_1,t}^{P2G_P}) + \frac{\rho_1}{2}(P_{j_1,t}^{P2G_G} - P_{j_1,t}^{P2G_P})^2 + \\ \sum_{j_2}\sum_t^{\varpi} \chi_{j_2,t}^G (G_{j_2,t}^{G2P_G} - G_{j_2,t}^{G2P_P}) + \frac{\rho_2}{2}(G_{j_2,t}^{G2P_G} - G_{j_2,t}^{G2P_P})^2 \\ s.t.\ (3)\text{-}(13), \forall t \in \mathbb{T} = \{1,2,\dots,\varpi\},\, j_1 \in \mathbb{J}_1,\, j_2 \in \mathbb{J}_2 \end{cases} \tag{31}$$

**Question 2:**

$$\begin{cases} \min L_2^E = -\ln(C_0^E - C^E) + \\ \sum_t^{\varpi} \gamma_t^P \left(\vartheta_t^{P2G_G} - \vartheta_t^{P2G_P}\right) + \frac{\rho_3}{2}(\vartheta_t^{P2G_G} - \vartheta_t^{P2G_P})^2 + \\ \sum_t^{\varpi} \gamma_t^G \left(\vartheta_t^{G2P_G} - \vartheta_t^{G2P_P}\right) + \frac{\rho_4}{2}(\vartheta_t^{G2P_G} - \vartheta_t^{G2P_P})^2 \\ s.t.\ (16)\text{-}(25), \forall t \in \mathbb{T} = \{1,2,\dots,\varpi\} \end{cases} \tag{32}$$

$$\begin{cases}\min L_2^G = -\ln(C_0^G - C^G) + \\ \sum_t^{\varpi} \gamma_t^P \left(\vartheta_t^{P2G_G} - \vartheta_t^{P2G_P}\right) + \frac{\rho_3}{2}\left(\vartheta_t^{P2G_G} - \vartheta_t^{P2G_P}\right)^2 + \\ \sum_t^{\varpi} \gamma_t^G \left(\vartheta_t^{G2P_G} - \vartheta_t^{G2P_P}\right) + \frac{\rho_4}{2}\left(\vartheta_t^{G2P_G} - \vartheta_t^{G2P_P}\right)^2 \\ s.t.\ (3)\text{-}(13),\ \forall t \in \mathbb{T} = \{1,2,\dots,\varpi\}\end{cases} \tag{33}$$

where $L_1^E$,$L_1^G$ ($L_2^E$,$L_2^G$) are the augmented Lagrangian functions for the ADN and GDN about **Question 1** (**Question 2**). $\chi_{j_1,t}^P$, $\chi_{j_2,t}^G$, $\gamma_t^P$, $\gamma_t^G$, $\rho_1$, $\rho_2$, $\rho_3$, $\rho_4$are the Lagrangian multipliers.

In summary, above Nash bargaining process can be understood as the social benefits obtained through jointly maximizing and then fairly sharing the social benefits. If any party submits false transaction information, it will cause the derived trading strategy to deviate from the solution that maximizes social welfare, ultimately reducing the distributed benefits [31]. Thus, ADN and GDN have no motivation to falsely report transaction information, which is in line with incentive compatibility.

*B. Two-stage Robust Nash Bargaining Its Solving Scheme*

Taking account of uncertainties, the ADN is at risk of failing to achieve the desired effect of prior trading decisions, therefore, a two-stage robust model is established, as shown in (34)-(35), where $\boldsymbol{y} = \left[P_{j_3,t}^{Li.pre}, P_{j_1,t}^{P2G}, P_{j_2,t}^{G2P}, \vartheta_t^{P2G}, \vartheta_t^{G2P}\right]$ is the first stage decision variable, $\boldsymbol{x} = \left[\Delta P_{j_3,t}^{Li}\right]$, is the second stage decision variable, $\boldsymbol{u} = \left[P_{j,t}^{LD.real}, P_{j_4,t}^{DER.real}\right]$. is the worst case of uncertain variable. Besides, based on constraints (16)-(25), the constraints of uncertain variables are added to (35).

$$\min_{\boldsymbol{y}} \max_{\boldsymbol{u}} \min_{\boldsymbol{x}} C^E \tag{34}$$

$$\begin{cases} s.t.\ (16)\text{-}(25) \\ P_{j_3,t}^{Li} = P_{j_3,t}^{Li.pre} + \Delta P_{j_3,t}^{Li} = P_{j_3,t}^{Li.disc} + P_{j_3,t}^{Li.c} \\ 0 \le (P_{j,t}^{LOAD.real} - P_{j,t}^{LOAD.pre})/P_{j,t}^{LOAD.pre} \le \phi^{LOAD} \\ 0 \le (P_{j_4,t}^{DER.real} - P_{j_4,t}^{DER.pre})/P_{j_4,t}^{DER.pre} \le \phi^{DER} \\ 0 \le P_{j_4,t}^{DER} \le P_{j_4,t}^{DER.real}, P_{j,t}^{LOAD} = P_{j,t}^{LOAD.real} \\ \forall t \in \mathbb{T} = \{1,2,\dots,\varpi\},\ j_1 \in \mathbb{J}_1,\ j_2 \in \mathbb{J}_2, \\ j_3 \in \mathbb{J}_3,\ j_4 \in \mathbb{J}_4,\ j \in \mathbb{B} \end{cases} \tag{35}$$

Through C&CG algorithm, the two-stage robust model is decomposed into the main problem (MP) and the subproblem (SP). As shown in Fig. 2, the MP serves to optimize the volume and prices of the energy transaction, through the model presented in section IV. A. The sub-problem is responsible for figuring out the worst cases of random variables, and planning the power adjustment profiles of li-ion batteries to de-risk profit collapse due to stochastic nature of DERs. Additionally, the detailed proof of the convergence of the algorithm proposed can be found in the Appendix.

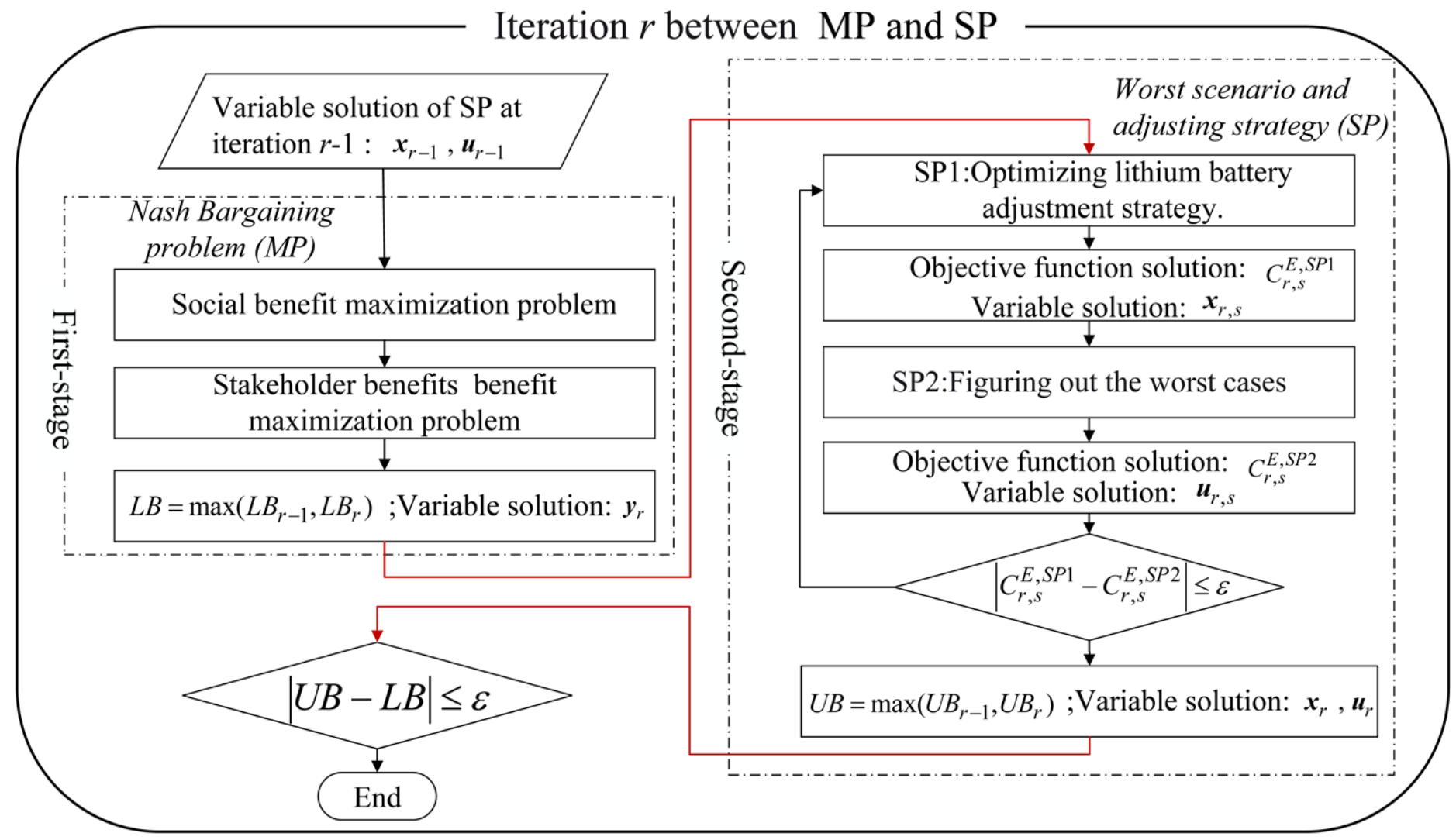

**Fig. 2.** Solving scheme of the proposed two-stage robust benefit sharing model between HCNG-enabled GDN and ADN

The MP and SP at iteration *r* of the C&CG algorithm is formulated as (36)-(37).

$$\text{MP:} \begin{cases} C_r^{E,SP} = \min_{\boldsymbol{y}} C^E \\ \boldsymbol{x} = \breve{\boldsymbol{x}}_{r-1}, \boldsymbol{u} = \breve{\boldsymbol{u}}_{r-1}, s.t.(26)\text{-}(33) \end{cases} \tag{36}$$

$$\text{SP:} \begin{cases} C_r^{E,SP} = \max_{\boldsymbol{u}} \min_{\boldsymbol{x}} C^E \\ \boldsymbol{y} = \breve{\boldsymbol{y}}_r, s.t.\ (16)\text{-}(25) \end{cases} \tag{37}$$

where $\check{\boldsymbol{x}}_{r-1}$, $\check{\boldsymbol{u}}_{r-1}$ are solutions of decision variable in the second stage and the worst case of SP, respectively, at iteration $r$-1, which will be set to a cutting plane constraint of MP. $\check{\boldsymbol{y}}_r$ is decision variable of MP solved in the first stage, at iteration $r$, which will be set to cutting plane constraints of SP. In this regard, MP and SP become a two-level solving loop.

Since the SP is a max-min problem and cannot be solved directly, to solve it, the BCD algorithm is used to decompose the SP into two tractable problems of subproblem1 (SP1) and subproblem2 (SP2). Note that, SP1 and SP2 construct an inner loop of their upstream solving loop of MP-SP.

In the at $s$th iteration of SP1-SP2 inner solving loop, the BCD algorithm sequentially solves (38) and (39).

$$\text{SP1:} \begin{cases} C_{r,s}^{E,SP1} = \min\limits_{\boldsymbol{x}} C^E \\ \boldsymbol{y} = \check{\boldsymbol{y}}_r,\ \boldsymbol{u} = \check{\boldsymbol{u}}_{r,s-1}, s.t. (16)\text{-}(25) \end{cases} \tag{38}$$

where the random variables $\boldsymbol{u}$ are fixed as $\check{\boldsymbol{u}}_{r,s-1}$, which is the worst case of SP2 at iteration $s-1$, during the $r$th solving loop of MP and SP.

$$\text{SP2:} \begin{cases} C_{r,s}^{E,SP2} = \max\limits_{\boldsymbol{u}} C^E \\ \boldsymbol{y} = \check{\boldsymbol{y}}_r, \boldsymbol{x} = \check{\boldsymbol{x}}_{r,s}, s.t. (16)\text{-}(25) \end{cases} \tag{39}$$

where $\boldsymbol{x}$ is fixed as $\check{\boldsymbol{x}}_{r,s}$, which is the optimal decision variable of SP1 at iteration $s$.

### *C. Convergence Proof of the Proposed Solving Algorithm*

To solve the proposed two-stage robust benefit sharing model between HCNG-enabled GDN and ADN, a new ADMM and BCD joint algorithm bridged by C&CG is proposed in this paper. The following theorem establishes the convergence for the ADMM and BCD joint algorithm bridged by C&CG.

**Theorem 1**: The proposed algorithm should satisfy the following three conditions for convergence:

1) Objective function$C^E$ and $C^G$in MP are closed and convex.

2) Objective function$C^E$ in SP is continuously differentiable and convex

3) Feasible region of variable $\boldsymbol{u}$ is a bounded closed set.

**Proof.** Please see Appendix.

## V. Neural Network Encoder for Time-Adaptive Two-stage Robust Nash Bargaining-based Energy Trading

The priori determination of optimal time window counts ($\varpi$) is complicated by uncertain trade-off between scheduling strategy quality and solution efficiency. To resolve this challenge, this paper develop a specialized neural network (NN) encoder architecture for selecting optimal $\varpi$.

### *A. Neural Network Encoder Training.*

#### *1) Training Sample Generation*

The training sample comprises (1) matrix $\boldsymbol{\zeta}$, representing the combined output of wind power, solar power and load, (2) vector $\boldsymbol{\varpi}$ with different assignments related to $\varpi$, (3) label set $\boldsymbol{\rho}$, representing scheduling strategy quality, (4) label set $\boldsymbol{\mathcal{T}}$, containing solving time of (36)-(39).

$$\boldsymbol{\varpi} = [\varpi_1, \dots \varpi_\kappa, \dots \varpi_{\mathrm{K}}] \in \mathbb{R}^{1\times \mathrm{K}} \tag{40}$$

$$\boldsymbol{\rho} = [\rho_1, \dots \rho_\kappa, \dots \rho_{\mathrm{K}}] \in \mathbb{R}^{1\times \mathrm{K}}, \tag{41}$$

$$\boldsymbol{\mathcal{T}} = [\mathcal{T}_1, \dots \mathcal{T}_\kappa, \dots \mathcal{T}_{\mathrm{K}}] \in \mathbb{R}^{1\times \mathrm{K}}, \tag{42}$$

The label set $\boldsymbol{\rho}$ and $\boldsymbol{\mathcal{T}}$ are generated according to (43)-(45). Firstly, for each input $\varpi_\kappa$, (43) solve the optimization problem defined by (36)-(39) to obtain solution matrix $\boldsymbol{v}_\kappa^*$ and record solving time $\mathcal{T}_\kappa$. $\boldsymbol{v}_\kappa^*$ contains the energy transaction volume and price components.

$$\begin{cases} \boldsymbol{v}_\kappa^* \triangleq \left[\boldsymbol{v}_{\kappa,1}^*, \dots, \boldsymbol{v}_{\kappa,t}^*, \dots, \boldsymbol{v}_{\kappa,\varpi_\kappa}^*\right] \xleftarrow{\text{solve}} (36)\sim(39) \leftarrow \varpi_\kappa, \boldsymbol{\zeta} \\ \boldsymbol{v}_{\kappa,t}^* = \left[\check{P}_{j_1,t}^{P2G}, \check{P}_{j_2,t}^{G2P}, \check{\vartheta}_t^{P2G}, \check{\vartheta}_t^{G2P}\right]^T \end{cases} \tag{43}$$

Secondly, (36)-(39) is solved with a fixed 1-hour time resolution to obtain solution vector $\boldsymbol{v}_R$, which also contains the energy trading volume and the energy trading price.

$$\begin{cases} \boldsymbol{v}_R \triangleq \left[\boldsymbol{v}_{R,1}, \dots, \boldsymbol{v}_{R,f}, \dots, \boldsymbol{v}_{R,24}\right] \xleftarrow{\text{solve}} (36)\sim(39) \leftarrow 24, \boldsymbol{\zeta} \\ \boldsymbol{v}_{R,f} = \left[\check{P}_{j_1,f}^{P2G}, \check{P}_{j_2,f}^{G2P}, \check{\vartheta}_f^{P2G}, \check{\vartheta}_f^{G2P}\right]^T \end{cases} \tag{44}$$

Since$\boldsymbol{v}_\kappa^*$ and $\boldsymbol{v}_R$ have temporal misalignment, (45) upsamples $\boldsymbol{v}^*$ by filling skipped time points with their assigned cluster centroids, producing the time-synchronized sequence $\boldsymbol{v}_{\kappa,R}^*$.

$$\begin{cases} \boldsymbol{v}_{\kappa,R}^* \triangleq \left[\boldsymbol{v}_{\kappa,R,1}^*, \dots, \boldsymbol{v}_{\kappa,R,f}^*, \dots, \boldsymbol{v}_{\kappa,R,24}^*\right] \\ \boldsymbol{v}_{\kappa,R,f}^* = \boldsymbol{v}_{\kappa,t}^*, \text{if } \sum_{u=1}^{t} N_u - N_t < f \leq \sum_{u=1}^{t} N_u \end{cases} \tag{45}$$

Finally, the variable $\left[P_{j_1,t}^{P2G}, P_{j_2,t}^{G2P}, \vartheta_t^{P2G}, \vartheta_t^{G2P}\right]$ in (36)-(39) is constrained to $\boldsymbol{v}_{\kappa,R}^*/\boldsymbol{v}_R$, yielding the target solution $C_{\kappa,R}^E, C_{\kappa,R}^G/\ C_R^E, C_R^G$.

Based on this, each label $\rho_\kappa$ is defined as the absolute social welfare difference:

$$\begin{cases} C^E_{\kappa,R}, C^G_{\kappa,R} \overset{\text{solve}}{\triangleq\!\longleftarrow} (36)\sim(39) \leftarrow 24, \boldsymbol{\zeta}, \boldsymbol{v}^*_{\kappa,R} \\ C^E_R, C^G_R \overset{\text{solve}}{\triangleq\!\longleftarrow} (36)\sim(39) \leftarrow 24, \boldsymbol{\zeta}, \boldsymbol{v}_R \\ \rho_\kappa = \left|(C^E_0 - C^E_{\kappa,R} + C^G_0 - C^G_{\kappa,R}) - (C^E_0 - C^E_R + C^G_0 - C^G_R)\right| \end{cases} \tag{46}$$

*2) Parameter Training in Neural Network*

Two fully connected neural networks $\Phi^{\boldsymbol{\theta}_\rho}$, $\Phi^{\boldsymbol{\theta}_\mathcal{T}}$ are trained on the sample set $\boldsymbol{\zeta}$, $\boldsymbol{\varpi}$, $\boldsymbol{\rho}$,$\boldsymbol{\mathcal{T}}$ as defined in (47)-(48):

$$\widetilde{\boldsymbol{\rho}} = \Phi^{\boldsymbol{\theta}_\rho}(\boldsymbol{\varpi},\boldsymbol{\zeta}), \boldsymbol{\theta}^*_\rho = \arg\min_{\boldsymbol{\theta}_\rho} \|\widetilde{\boldsymbol{\rho}} - \boldsymbol{\rho}\|^2_2 \tag{47}$$

$$\widetilde{\boldsymbol{\mathcal{T}}} = \Phi^{\boldsymbol{\theta}_\mathcal{T}}(\boldsymbol{\varpi},\boldsymbol{\zeta}), \boldsymbol{\theta}^*_\mathcal{T} = \arg\min_{\boldsymbol{\theta}_\mathcal{T}} \left\|\widetilde{\boldsymbol{\mathcal{T}}} - \boldsymbol{\mathcal{T}}\right\|^2_2 \tag{48}$$

where structural parameters $\boldsymbol{\theta}_\mathcal{T}$ and $\boldsymbol{\theta}_\rho$ govern the network's estimation of solution time and scheduling quality, producing outputs $\widetilde{\boldsymbol{\rho}}$ and $\widetilde{\boldsymbol{\mathcal{T}}}$. $\boldsymbol{\theta}^*_\rho$ and $\boldsymbol{\theta}^*_\mathcal{T}$ represent the optimally trained structural parameters.

*B. Optimal $\varpi$ Selection Based on Neural Network Encoder*

Given the pre-trained neural network $\Phi^{\boldsymbol{\theta}^*_\rho}$ and $\Phi^{\boldsymbol{\theta}^*_\mathcal{T}}$, the optimal parameter $\varpi$ in models (36)-(39) is efficiently computed via (49), ensuring compliance with the prescribed constraint of solving time $\epsilon^\mathcal{T}$.

$$\begin{cases} \widetilde{\varpi}^* = \arg\min_{\varpi} \tilde{\rho} = \Phi^{\boldsymbol{\theta}^*_\rho}(\varpi, \boldsymbol{\zeta}) \\ s.t. \tilde{\mathcal{T}} = \Phi^{\boldsymbol{\theta}^*_\mathcal{T}}(\varpi, \boldsymbol{\zeta}) \le \epsilon^\mathcal{T} \end{cases} \tag{49}$$

## VI. Numerical Studies

A IENGS consisting of an IEEE 33-bus ADN and a 20-node Belgian GDN is studied, as shown in Fig.3, where solar and wind power generation data are from Elia, power and gas load data are collected from [32].

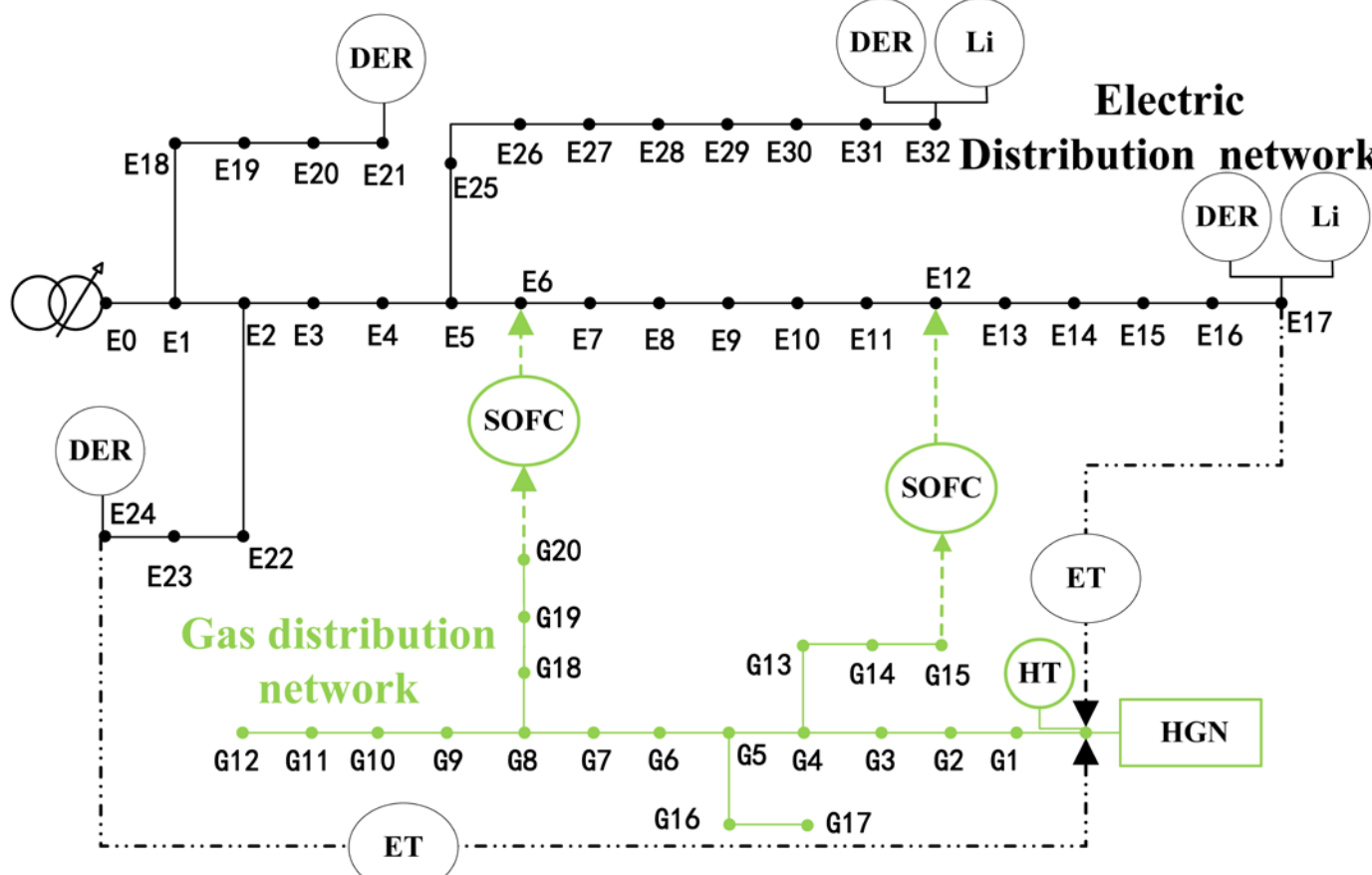

**Fig. 3.** Configuration diagram of combined optimization model between electricity and HCNG distribution networks.

In IEEE 33-bus ADN, DERs are integrated at bus E17, E21, E24, and E32, bus E17 and E32 own li-ion batteries, bus E6 and E12 are connected to node G8 and G4 of the GDN through SOFCs. Voltage limits of all power buses are within [0.95, 1.05] p.u. The uncertainties of DER generation and load are simulated via 15% and 5% intervals upon their forecasting baselines. In 20-node GDN, HT is integrated at node G0, and two ETs are set on node E17 and E24.

*A. Efficiency analysis of cooperative ADN and GDN*

To prove the effectiveness of the proposed cooperative operational framework, three competitive models are given:

**Model 1**: The proposed cooperative ADN and GDN.

**Model 2**: No cooperation between ADN and GDN. Following [33], ADN configures its own li-ion batteries, HT and SOFCs, and goes self-energy conversion between electricity and hydrogen. The **Model 2** can be deemed as a development of the architecture in [34], and it can help figure out if SOFCs can turn this self-energy conversion loop to be profitable.

**Model 3**: No cooperation between ADN and GDN, and only li-ion batteries are included in ADN [35]. This is a common case known as active distribution network. The **Model 3** comes out to highlight the high capacity and cost-effectiveness merits of the proposed IENGS architecture.

Without uncertainty involved, the daily operating costs, energy conversion loss rate, and the marginal cost of ET and SOFC (i.e., the sensitivity of their cost versus the produced energy) are shown in Tab. I. Benefit ingredients and energy conversion profiles are shown in Tab. II and Fig. 4.

TABLE I
Daily Operating Costs Without Uncertainty Considered

| Model # | Operating Costs $ | | Energy Conversion Loss Rate | Marginal cost $/kWh |
|---|---|---|---|---|
| **Model 1** | ADN | **1537** | **27%** | **0.0307** |
| | GDN | **23614** | | |
| **Model 2** | ADN | 1699 | 44% | 0.0719 |
| | GDN | 23809 | | |
| **Model 3** | ADN | 1755 | - | - |
| | GDN | 23809 | | |

TABLE II
INGREDIENTS ANALYSIS OF COST AND BENEFIT ($)

| Model | | Benefit* | $C^{SOFC}$ | $C^{ET}$ | $C^{TG}$ | $C^{HGN}$ |
|---|---|---|---|---|---|---|
| **Model 1** | ADN | 218 | 273 | 116 | 240 | 24220 |
| | GDN | 194 | | | | |
| **Model 2** | ADN | 56 | 27 | 34 | 1271 | 23809 |

*Benefits of ADN and GDN are calculated via $C_0^E - C^E$ and $C_0^G - C^G$, respectively.

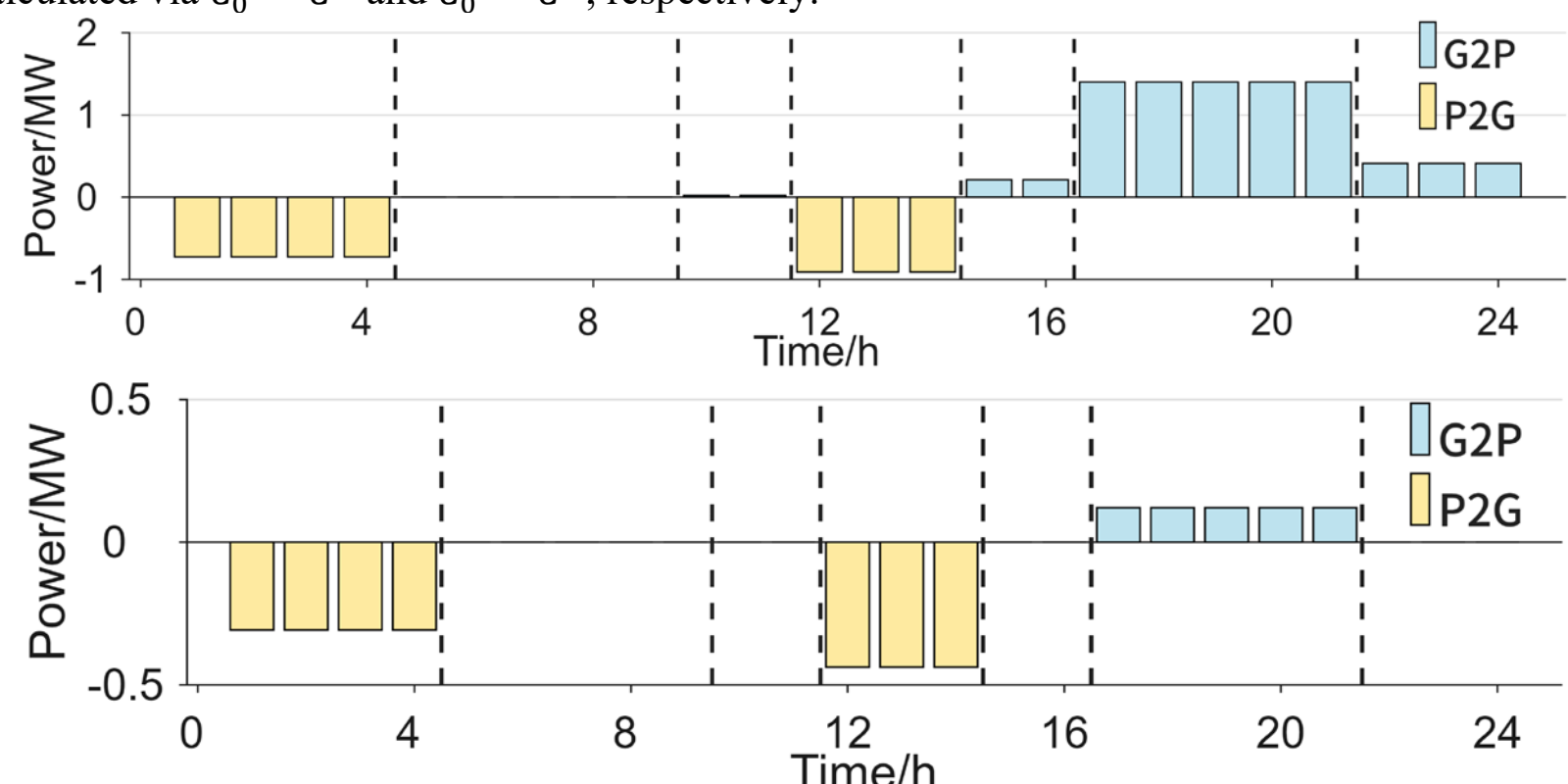

**Fig. 4.** Energy conversion profiles. Note that, positive power indicates that the energy conversion runs from gas to power, also the power injection to the ADN (top: **Model 1**. bottom: **Model 2**; The dashed line demarcates the time-adaptive window boundaries).

From Tab. I and Fig. 4, daily operating cost of ADN by **Model 1** is similar to that by **Model 2**, both reducing the ADN's cost by nearly 12% against **Model 3**. However, **Model 1** significantly outperform **Model 2** regarding daily operating costs of GDN. **Model 1** creates the best social welfare which is nearly 1.6% to total cost, while guaranteeing ADN benefits. In **Model 1**, the ADN profits by selling hydrogen or purchasing low-cost HCNG for power generation, and the GDN gains by selling natural gas and producing low-priced hydrogen. In contrast, **Model 2** relies solely on the price arbitrage between local renewables and the upstream grid. Furthermore, **Model 2**'s restriction to a pure hydrogen-power conversion cycle, without the GDN's fuel support, leads to a high energy loss rate of 44% and an inferior marginal conversion cost ($0.0719/kWh vs. $0.0307/kWh in **Model 1**), indicating significant energy waste and diminished social welfare. Additionally, Tab. II confirms that **Model 1** successfully distributes the generated benefits between the ADN and GDN via the Nash bargaining model, validating the proposed sharing strategy.

*B. Verification of the proposed robust benefit sharing strategy*

Robust benefit sharing de-risks profit collapse due to forecasting error. Here we consider three robust cases (**Case 2**~**Case 4**). Notably, **Case 1** is a deterministic case that assumes forecasts are exact, which in fact corresponds to single MP.

**Case 1**: The forecasts of load and DERs generation are exact.

**Case 2**: The forecast of DER generation is exact, while load profile is produced by the robust optimization procedure SP2.

**Case 3**: The forecast of load is exact, DER generation profile is produced by the robust optimization procedure SP2.

**Case 4**: Load and DERs generation profiles are both produced by the robust optimization procedure SP2.

**Case 2**~**Case 4** establish the testing robust operational set. To show the importance of li-ion batteries in robust benefit sharing, daily operating costs with or without lithium battery power redispatch as shown in Tab. III.

TABLE III
OPERATING COSTS WITH OR WITHOUT BATTERY POWER REDISPATCH ($)

| Entities | **Case 1** | **Case 2** | | **Case 3** | | **Case 4** | |
|---|---|---|---|---|---|---|---|
| | | No bat. | With bat. | No bat. | With bat. | No bat. | With bat. |
| ADN | 1537 | 2017 | 1922 | 3038 | 2859 | 3636 | 3459 |
| GDN | 23614 | 23614 | | 23614 | | 23614 | |

Tab. III reveals two key findings. First, as expected, operating costs rise when robustness is considered. Second, and more importantly, under **Case 2**~**Case 4**, li-ion batteries improve operational economy by an average of 5.5%. This occurs because SOFCs and ETs cannot respond rapidly enough to balance real-time fluctuations from DERs and load. Consequently, the ADN must rely on either fast-regulation assets or costly power purchases from the upstream grid. Li-ion batteries mitigate this by reducing expensive grid purchases, offsetting their own lifetime costs in the process.

We further analyze how robust benefit sharing mitigates profit collapse for the ADN (Tab. IV). Forecasting errors inevitably cause intraday power imbalances. To maintain balance, the ADN incurs costs, whether for battery auxiliary services or grid power. Its profit collapse stems from the transaction bias between the day-ahead cooperative schedule and the actual intraday operations. So, a trivial assumption that only considers electricity purchase from upstream power grid is made during the forming of Tab. IV.

TABLE IV
COSTS OF DAY-AHEAD SCHEDULES ON INTRADAY CASES ($)

| Testing intraday operational case/set | Proposed Robust benefit sharing (with SP) | | Deterministic benefit sharing (excluding SP) | |
|---|---|---|---|---|
| | $C^E$ | $C^{TG}$ | $C^E$ | $C^{TG}$ |
| **Case 1** | 1537 | 240 | 1537 | 240 |
| **Case 2** | 1906 | 399 | 1922 | 857 |
| **Case 3** | 2551 | 776 | 2859 | 1919 |
| **Case 4** | 3089 | 1250 | 3459 | 2519 |
| *Mean±var.* stat. under robust set | 2271±595 | 666±389 | 2444±758 | 1309±892 |

In. Tab. IV, there is clear trends of decreasing operating cost of the ADN ($C^E$) by the proposed method. Also, more uncertainties enlarge the risk of profit collapse, as you can see **Case 4** manifests the largest increase of cost in the comparison between our method and the deterministic one. The cost statistics revealed that, the robust strategy indeed stabilizes the profit of the ADN to withstand forecasting errors of uncertainties, and its decision pattern is to facilitate interactions between the ADN and the GDN to reduce costly electricity purchase from upstream power grid ($C^{TG}$), which will be further analyzed later.

To further justify our framework, Tab. V, Tab. VI, Fig. 5, and Fig. 6 are provided. These outcomes stem from comparative experiments under the given robust operating cases.

TABLE V
TOTAL OPERATING COSTS BY COMPETITIVE MODELS IN DIFFERENT ROBUST CASES ($)

| Case # | Entities | **Model 1** | **Model 2** | **Model 3** |
|---|---|---|---|---|
| **Case 2** | ADN | **1906** | 2011 | 2143 |
| | GDN | **23579** | 23809 | 23809 |
| **Case 3** | ADN | **2551** | 2816 | 2820 |
| | GDN | **23558** | 23809 | 23809 |
| **Case 4** | ADN | **3089** | 3382 | 3382 |
| | GDN | **23480** | 23809 | 23809 |

TABLE VI
BENEFIT ANALYSIS IN DIFFERENT ROBUST CASES ($)

| Case # | Entities | **Model 1** | **Model 2** |
|---|---|---|---|
| **Case 2** | ADN | 237 | 56 |
| | GDN | 230 | 0 |
| | Total | 467 | 56 |
| **Case 3** | ADN | 269 | 4 |
| | GDN | 251 | 0 |
| | Total | 520 | 4 |
| **Case 4** | ADN | 293 | |
| | GDN | 329 | 0 |
| | Total | 622 | |

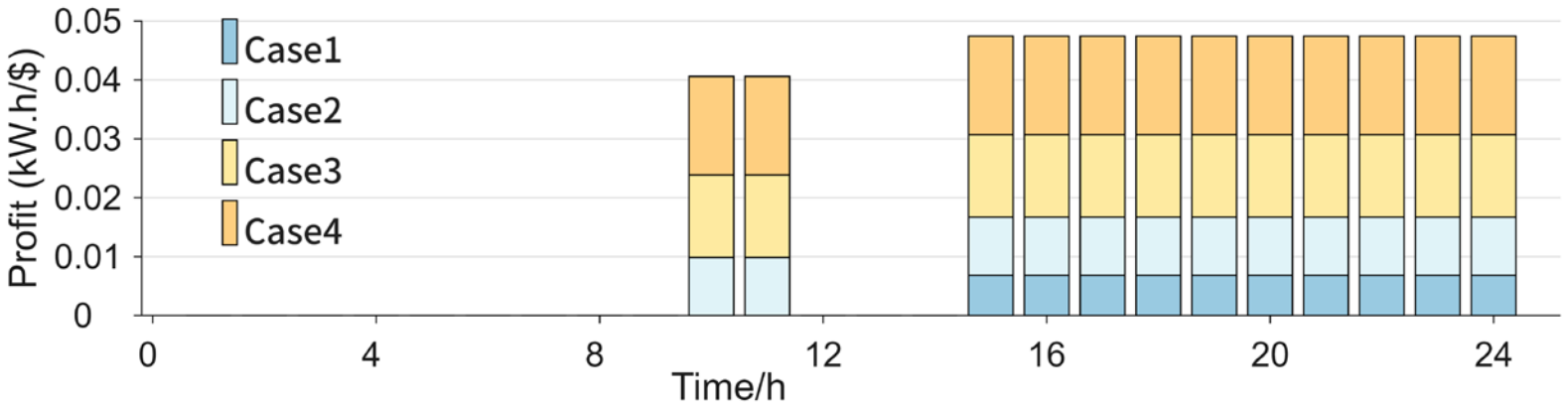

**Fig. 5.** Marginal profit of the GDN selling HCNG upon the proposed benefit sharing mechanism (**Model 1**) under the testing robust cases.

As shown in Tabs. V and VI, the proposed method (**Model 1**) achieves superior cost-benefit stability compared to its rivals. This is exemplified by the sharp benefit decline of **Model** from **Case 2** to **Case 3**, whereas our method maintains stable benefits for both the ADN and GDN across all extreme cases. Moreover, **Model 1** yields the best benefit for the GDN, which gains extra income through energy conversion with the ADN. This phenomenon is further elucidated in Fig. 5: the marginal profit for the GDN selling HCNG rises from **Case 1** to the worst **Case 4**. More severe operating conditions for the ADN incentivize increased HCNG purchases, boosting GDN revenue, which is then shared back with the ADN via the cooperative mechanism, ensuring mutually stable and risk-averse benefits. In this win-win situation, the GDN secures extra payback while helping stabilize the ADN, thereby consolidating long-term gains.

Fig. 6 underscores the significance of the SOFC-bridged cooperative mechanism. During hours 15~24, when DER generation plummets, G2P transactions in **Model 1** are actively stimulated for two reasons. First, within the proposed IENGS, the marginal cost of HCNG-to-power remains stable and is significantly lower than that of electricity from the upstream grid. Second, while li-ion batteries offer the best marginal cost, they lack the sufficient controllable capacity to serve as critical flexible resources in extreme

**Case 4**, unlike the GDN's immense storage. Regarding **Model 2**, self-hydrogen storage manifests the largest marginal cost, thus the ADN would rather adopt electricity from upstream power grid than hydrogen storage. However, the former is noticeably costly than the applied HCNG-to-power (by over 20%), wherefore the proposed **Model 1** is more lucrative for growth.

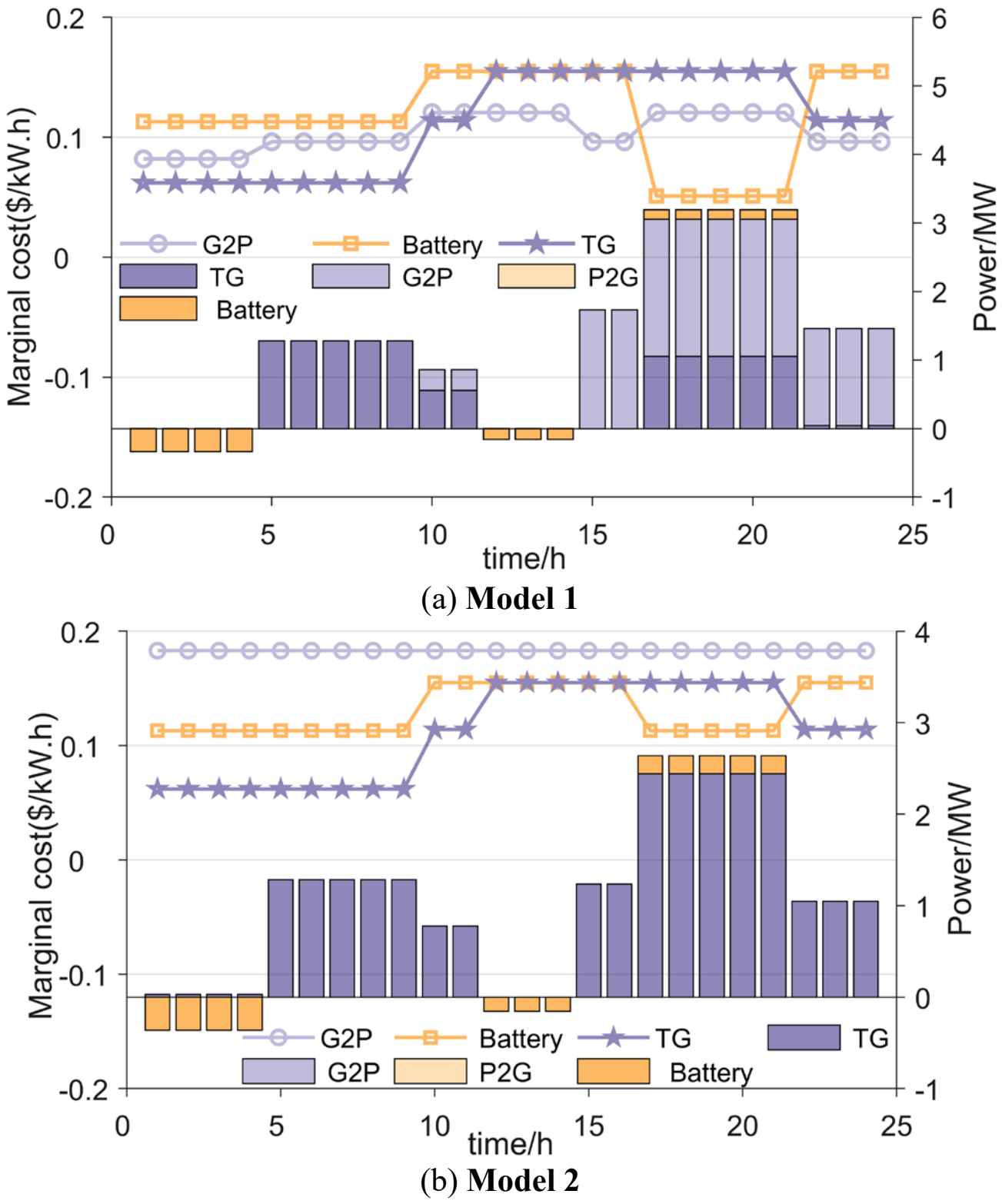

**Fig. 6.** Profiles of operations and marginal costs of involved infrastructures under **Case 4**. (left axis: line chart; right axis: bar chart)

### *C. Performance assessment of neural-enhanced time-adaptive robust Nash bargaining strategy*

Two comprehensive simulations are designed to quantitatively demonstrate the performance advantages of our model:

**S1**: The proposed neural-enhanced time-adaptive robust Nash bargaining strategy.

**S2**: Conventional scheduling strategy, which employs a rigid one-hour time resolutions.

TABLE VII
PERFORMANCE COMPARISON BETWEEN **S1** TO **S2**

| | | | | MP **(Question 1)** | MP **(Question 2)** | SP |
|---|---|---|---|---|---|---|
| | | | | ADMM | ADMM | BCD |
| **S1** | Time (s) | C&CG between the MP and SP | The first iteration | 35.74 | 25.10 | 7.64 |
| | | | The second iteration | 20.85 | 6.28 | 6.86 |
| | Total cost ($) | | | 25151 | | |
| **S2** | Time (s) | C&CG between the MP and SP | The first iteration | 134.76 | 66.36 | 22.72 |
| | | | The second iteration | 78.61 | 16.56 | 20.94 |
| | Total cost ($) | | | 25131 | | |

As evidenced by Table VII, our neural-enhanced time-adaptive robust Nash bargaining strategy achieves significant convergence acceleration across all model components. Most notably, the MP demonstrates a 99s reduction (73.48% efficiency gain) in Nash bargaining operations. While SP's relatively simpler architecture yields more modest absolute time savings (~12s), its 67% efficiency improvement remains substantial. Overall, the proposed strategy realizes a 237.48s total speedup with 69% average efficiency enhancement, all while with negligible impact on total cost (deviation < 0.08%).

## VII. CONCLUSION

This paper proposes a neural-enhanced, time-adaptive robust Nash bargaining strategy to enable stable cooperation between HCNG network and ADN, interconnected via SOFCs. The framework first employs Nash bargaining to actualize the social welfare potential of the efficient HCNG-SOFC energy conversion. To address convergence challenges and maintain bilateral benefits, a neural network is introduced to govern adaptive time windows. Finally, the incorporation of li-ion batteries within a two-stage robust model de-risks potential profit collapse for the ADN. Based on the time-adaptive window, the first stage ensures privacy of the two utilities via ADMM, the second stage efficiently dispatches li-ion batteries to withstand the worst operational cases of ADN via emerging BCD algorithm, and the two stages are looped upon C&CG. Numerical study shows that the proposed model circumvents profit collapse, fulfills stable gains for both GDN and ADN, and reaches a stable social welfare of nearly 1.6% to total cost and improve computational efficiency by over 69.86% versus the fixed time-scale baseline.

## APPENDIX

The proof of **Theorem 1** consists of three main parts, as detailed below:

### *A. The convergence of ADMM for MP*

**Lemma 1.** [36]: The ADMM algorithm solves problems in the form: $\min f(x) + g(x)$, if. the (extended-real-valued) functions $f: \mathrm{R}^n \rightarrow \mathrm{R} \cup \{+\infty\}$ and $g: \mathrm{R}^m \rightarrow \mathrm{R} \cup \{+\infty\}$ are closed and convex, ADMM algorithm is converge.

**Proof.** In section IV(A), Nash bargaining model is transformed into **Question 1** and **Question 2**. The transaction price variables ($[\vartheta_t^{P2G}, \vartheta_t^{G2P}]$) in **Question 1** are eliminated, thus, the objective function in (27) and (30)-(31) are linear functions about variables: $\{P_{j_1,t}^{P2G}, G_{j_2,t}^{G2P}, G_{j_1,t}^{HT}, P_{j_3,t}^{Li}, P_{j_1,t}^{P2G}, P_{j_2,t}^{G2P}\}$, where the objective function in (27) and (30)-(31) can be considered $\min f(x) + g(x)$, $f$ and $g$ in Lemma 1. Therefore, the objective functions and the feasible region in (30)-(31) are convex and closed obviously and ADMM is converge in **Question 1**.

In **Question 2** energy transaction variables ($[P_{j_1,t}^{P2G}, G_{j_2,t}^{G2P}]$) multiplied by transaction price variables ($[\vartheta_t^{P2G}, \vartheta_t^{G2P}]$) are replaced with fixed parameter $[\check{P}_{j_1,t}^{P2G}, \check{G}_{j_2,t}^{G2P}]$. Thus, the objective function about variables $\{\vartheta_t^{P2G}, \vartheta_t^{G2P}\}$ in (28) and (32)-(33) are also convex. The next proof of ADMM's convergence in **Question 2** is same as **Question 1**. Due to the convergence of **Question 1** and **Question 2**, MP also converges. ☐

### *B. The convergence of BCD for SP*

**Lemma 2.** [37]: If the function $f$ in problem $\min f(\boldsymbol{x}): \boldsymbol{x} \in \mathrm{R}^n$ is continuously differentiable and convex, the optimal solution of function $\arg \min_{x_i \in \mathrm{R}^n} f(x_1, \dots, x_{i-1}, x_i, x_{i+1}, \dots, x_m)$ is unique and the BCD algorithm is converge.

**Proof.** The variables ($\boldsymbol{y} = [P_{j_3,t}^{Li.pre}, P_{j_1,t}^{P2G}, P_{j_2,t}^{G2P}, \vartheta_t^{P2G}, \vartheta_t^{G2P}]$) in SP(37) is replaced with fixed parameter $\breve{\boldsymbol{y}}_r$, thus, the objective function ($C^E$) of (37) is a linear function, which is continuously differentiable and convex obviously. Therefore, the SP in section IV(B) satisfies the convergence condition of *Lemma 2*

### *C. The convergence of the proposed ADMM and BCD joint algorithm bridged by C&CG*

To start our proof, we firstly introduce a necessary lemma regarding the convergence of the C&CG algorithm:

**Lemma 3.** [38,39]: The C&CG algorithm decomposes the original problem into main problems and sub-problems that iteratively interact with each other, and two sequences of upper and lower bounds of original problem $\{UB^t\}$, $\{LB^t\}$ are obtained during the process of algorithm iteration. If there exist two non-empty and bounded sequences $\{UB^t\}$, $\{LB^t\}$ such that $|UB^t - LB^t| \rightarrow 0$, C&CG algorithm will terminate in a finite number of iterations, then the C&CG algorithm is converge.

**Proof.** The optimal solution of the two-stage robust optimization is the objective value in the worst-case scenario, in other words, the optimal decision holds for all values of uncertainty scenarios ($\boldsymbol{u}$). The cutting planes added in the master problem only optimize under partial uncertainty scenarios, thus, MP (36) is a relaxation problem of original problem (34), and since it seeks to minimize $C^E$, $C_r^{E,MP}$ is definitely a lower bound for the original problem (34).

In addition, in SP (37), $\boldsymbol{y}$ is fixed to a constant value $\breve{\boldsymbol{y}}_r$ from (34), which is not necessarily the optimal solution of original problem (34). Thus, $C_r^{E,SP}$ is a feasible solution of (34), which is greater than or equal to the optimal solution, $C_r^{E,SP}$ is definitely an upper bound for the original problem (34).

In the above iteration process, both MP (34) and SP (37) search for the maximum or minimum value of $C^E$ within the feasible domain along the gradient, and obtain two sequences of upper and lower bounds ($C_r^{E,SP} \in \boldsymbol{UB}$ and $C_r^{E,MP} \in \boldsymbol{LB}$).

In the function $C^E$ of SP2 (39), $[\boldsymbol{y}, \boldsymbol{x}]$ are fixed as constants$[\breve{\boldsymbol{y}}_r, \breve{\boldsymbol{x}}_{r,s}]$, thus, only $C^{TG}$ include variables:

$$C^{TG} = \sum_t \mu_t P_t^{TG} \tag{A1}$$

$$P_t^{TG} = \sum_j (P_{j,t}^{LOAD} - P_{j,t}^{DER} - P_{j,t}^{G2P} - P_{j,t}^{Li} + P_{j,t}^{P2G}) \tag{A2}$$

where $[P_{j,t}^{G2P}, P_{j,t}^{Li}, P_{j,t}^{P2G}] \in [\boldsymbol{y}, \boldsymbol{x}]$ are fixed as constants, $P_{j,t}^{DER}$ and $P_{j,t}^{LOAD}$ are variables.

$$0 \leq P_{j_4,t}^{DER} \leq P_{j_4,t}^{DER.real}, P_{j,t}^{LOAD} = P_{j,t}^{LOAD.real} \tag{A3}$$

Taking into account the inequality constraint relationship between variable $P_{j_4,t}^{DER.real} \in \boldsymbol{u}$ and variable $P_{j_4,t}^{DER}$ in the objective function $C^E$, a set of variables $\boldsymbol{\kappa}$ are introduced and $\forall \kappa_{j_4,t} \in \boldsymbol{\kappa}$ satisfies the constraint of (A3):

$$\kappa_{j_4,t} + P_{j,t}^{DER} = P_{j_4,t}^{DER.real}, \kappa_{j_4,t}^{min} \leq \kappa_{j_4,t} \leq \kappa_{j_4,t}^{max} \tag{A4}$$

Based on the above deduction, the first-order Taylor expansion formula for variable $\boldsymbol{u} = [P_{j,t}^{LD.real}, P_{j_4,t}^{DER.real}]$ at point $\breve{\boldsymbol{u}}_{r,s-1} = [\check{P}_{j,t,r,s-1}^{LOAD}, \check{P}_{j_4,t,r,s-1}^{DER}]$ in $C^E$ of (38) is:

$$F(\boldsymbol{u}) = C_{r,s-1}^{E,SP2} + \sum_t \sum_j \mu_t (P_{j,t}^{LOAD.real} - \check{P}_{j,t,r,s-1}^{LOAD}) + \sum_t \sum_{j_4} (\mu_t (\kappa_{j_4,t} - \check{\kappa}_{j_4,t,r,s-1}) - \mu_t (P_{j_4,t}^{DER.real} - \check{P}_{j_4,t,r,s-1}^{DER})) \tag{A5}$$

In order to maximize the value of $C^E$, the variable $\kappa_{j_4,t}$/ $P_{j_4,t}^{LOAD.real}$ tends to approach the maximum value $\kappa_{j_4,t}^{max}$/ $P_{j_4}^{LOAD.max}$ within its range, the variable $P_{j_4,t}^{DER.real}$ tends to approach the maximum value $P_{j_4}^{DER.min}$ within its range. If the number of iterations between SP1 and SP2 tends to infinity ($s \rightarrow \infty$), the target solution $C_r^{E,SP}$ and variable solution $\breve{\boldsymbol{u}}_r$ ate obtained, and if the number of iterations between MP and SP tends to infinity ($r \rightarrow \infty$), $\kappa_{j_4,t}$/ $P_{j_4,t}^{LOAD.real}$ /$P_{j_4,t}^{DER.real}$ will converge to $a_{\kappa}/a_{load}/a_{DER}$ ($a_{\kappa} \leq \kappa_{j_4,t}^{max}$, $a_{load} \leq P_{j_4}^{LOAD.max}$, $a_{DER} \leq P_{j_4,t}^{DER.real}$), then, $|\breve{\boldsymbol{u}}_r - \breve{\boldsymbol{u}}_{r-1}| \rightarrow 0$.

The optimal values of both the first-stage and second-stage variables $[\boldsymbol{y}, \boldsymbol{x}]$ change due to variations in the scenario variables $\boldsymbol{u}$. If $|\breve{\boldsymbol{u}}_r - \breve{\boldsymbol{u}}_{r-1}| \rightarrow 0$, variables $\boldsymbol{y}$ and $\boldsymbol{x}$ will trend towards convergence as well, then, $\left|C_r^{E,SP} - C_{r-1}^{E,MP}\right| \rightarrow 0$.

Finally, based on the proof in Appendix A, MP and SP converge separately. Furthermore, because the feasible region of variable $\boldsymbol{u}$ is bounded and closed, the number of iterations r between the main problem and the subproblem is finite. Thus, there must exist two bounded sequences ($\boldsymbol{UB}$ and $\boldsymbol{LB}$) for that make the proposed ADMM and BCD joint algorithm bridged by C&CG converge in a finite number of iterations. ☐